\documentclass[11pt]{article}
\usepackage{amsmath, amssymb, amsfonts, amsthm}

\date{}
\newtheorem{theorem}{Theorem}[section]
\newtheorem{proposition}[theorem]{Proposition}

\def\E^#1{{\buildrel #1 \over\vee}}

\newcommand{\be}{\begin{equation}}
\newcommand{\ee}{\end{equation}}
\newcommand{\bey}{\begin{eqnarray}}
\newcommand{\eey}{\end{eqnarray}}

\input epsf

\newcommand{\donothing}[1]{}

\begin{document}
\title{Groups of Operators for Evolution Equations \\ of Quantum Many-Particle Systems}
\author{V.I. Gerasimenko\thanks{Partially supported by the WTZ grant No M/124 (UA 04/2007)}\\
\\
Institute of Mathematics of NAS of Ukraine,
          Tereshchenkivs'ka Str., 3,\\ 01601 Kyiv-4, Ukraine\\
gerasym@imath.kiev.ua \\}

\maketitle

\begin{abstract}
The aim of this work is to study the properties of groups of operators for evolution equations
of quantum many-particle systems, namely, the von Neumann hierarchy
for correlation operators, the BBGKY hierarchy for marginal density operators and
the dual BBGKY hierarchy for marginal observables. We show that
the concept of cumulants (semi-invariants) of groups of operators for the von Neumann equations
forms the basis of the expansions for one-parametric families of operators for
evolution equations of infinitely many particles.
\end{abstract}

\author{MSC(2000): 35Q40; 47d06.cumulant (semi-invariant),
           BBGKY hierarchy, dual BBGKY hierarchy, von Neumann hierarchy.}

\section{Introduction}
\setcounter{equation}{0}

Recently we observe significant progress in the study of the evolution equations of quantum many-particle systems \cite{CGP97},\cite{ AA}.
In particular it is involved in such fundamental problem as the rigorous
derivation of quantum kinetic equations \cite{Sp07}-\cite{FL}.

As is well known, there are various possibilities to describe the evolution of quantum many-particle systems \cite{CGP97}.
The sequence of the von Neumann equations for density operators \cite{AA},\cite{ Pe95},  the von Neumann hierarchy
for correlation operators \cite{GerSh}, the BBGKY hierarchy for marginal density operators \cite{Pe95} and
the dual BBGKY hierarchy for marginal observables \cite{CGP97} give the equivalent approaches for
the description of the evolution of finitely many particles. Papers \cite{GerSh}-\cite{GerRS}
constructed the one-parametric families of operators that define solutions of
the Cauchy problem for these evolution equations. It was established that
the concept of cumulants (semi-invariants) of groups of operators for the von Neumann equations
forms the basis of the one-parametric families of operators of various
evolution equations of quantum systems of particles, in particular, the BBGKY hierarchy
for infinitely many particles \cite{GerSh}.

The aim of the paper is to investigate properties of groups of operators for evolution equations
of quantum many-particle systems related with their cumulant structure on suitable Banach spaces.

In the beginning we will formulate some necessary facts about the description of quantum many-particle systems.

\subsection{Group of Operators for the von Neumann Equation}
Let a sequence $f=\big(I,f_{1},\ldots,f_{n},\ldots\big)$ is an infinite sequence of self-adjoint
operators $f_{n}$ ($I$ is a unit operator) defined on the Fock space
$\mathcal{F}_{\mathcal{H}}={\bigoplus\limits}_{n=0}^{\infty}\mathcal{H}^{\otimes n}$ over the Hilbert space
$\mathcal{H}$~ ($\mathcal{H}^{0}=\mathbb{C}$).
Operators $f_{n}$ defined in the $n$-particle
Hilbert space $\mathcal{H}_{n}=\mathcal{H}^{\otimes
n}$ we will denote by $f_{n}(1,\ldots,n)$.
For a system of identical particles obeying
Maxwell-Boltzmann statistics, one has $f_{n}(1,\ldots,n)=f_{n}(i_1,\ldots,i_n)$ if $\{i_{1},\ldots,i_{n}\}\in \{1,\ldots,n\}$.

Let $\mathfrak{L}^{1}_{\alpha}(\mathcal{F}_\mathcal{H})= {\bigoplus\limits}_{n=0}^{\infty}
\alpha^{n}\mathfrak{L}^{1}(\mathcal{H}_{n})$ be the space of sequences
$f=\big(I,f_{1},\ldots,f_{n},\ldots\big)$ of trace class operators
$f_{n}=f_{n}(1,\ldots,n)\in\mathfrak{L}^{1}(\mathcal{H}_{n})$, satisfying the above-mentioned  symmetry condition,
equipped with the trace norm
\begin{eqnarray*}
            \|f\|_{\mathfrak{L}^{1}_{\alpha} (\mathcal{F}_\mathcal{H})}=
            \sum\limits_{n=0}^{\infty}~ \alpha^{n} \|f_{n}\|_{\mathfrak{L}^{1}(\mathcal{H}_{n})}=
            \sum\limits_{n=0}^{\infty}~ \alpha^{n}~\text{Tr}_{\mathrm{1,\ldots,n}}|f_{n}(1,\ldots,n)|,
\end{eqnarray*}
where $\alpha>1$ is a real number, $\text{Tr}_{\mathrm{1,\ldots,n}}$ is the partial trace over $1,\ldots,n$ particles.
We will denote by $\mathfrak{L}^{1}_{\alpha, 0}$ the everywhere dense set
in $\mathfrak{L}^{1}_{\alpha}(\mathcal{F}_\mathcal{H})$ of finite sequences
of degenerate operators \cite{Kato} with infinitely differentiable kernels with compact supports.
We will also consider the space $\mathfrak{L}^{1}(\mathcal{F}_\mathcal{H})= {\bigoplus\limits}_{n=0}^{\infty}
\mathfrak{L}^{1}(\mathcal{H}_{n})$.

We note that the sequences of operators $f_{n}\in\mathfrak{L}^{1}(\mathcal{H}_{n}),$ $n\geq 1$, whose kernels are known
as density matrices \cite{BerSh} defined on the $n$-particle
Hilbert space $\mathcal{H}_{n}=\mathcal{H}^{\otimes
n}=L^{2}(\mathbb{R}^{\nu n})$, describe the states of a quantum system of non-fixed number of particles.
The space
$\mathfrak{L}^{1}(\mathcal{F}_\mathcal{H})$ contains sequences of operators more general than those determining
the states of systems.

The evolution of all possible states of quantum systems is described by the initial-value problem
to the von Neumann equation \cite{DauL_5, Pe95}.
A solution of such Cauchy problem is defined by the following one-parametric family of operators
on $\mathfrak{L}^{1}(\mathcal{F}_\mathcal{H})$
\begin{eqnarray}\label{groupG}
\mathbb{R}^{1}\ni t\mapsto\mathcal{G}(-t)f:= \mathcal{U}(-t)f\mathcal{U}^{-1}(-t),
\end{eqnarray}
where $f\in\mathfrak{L}^{1}(\mathcal{F}_\mathcal{H})$ and~~ $\mathcal{U}(-t)={\bigoplus\limits}_{n=0}^{\infty}\mathcal{U}_{n}(-t)$,
\begin{eqnarray}\label{evol_oper}
                &&\mathcal{U}_{n}(-t):=e^{-{\frac{i}{\hbar}}tH_{n}},\nonumber\\
                &&\mathcal{U}_{n}^{-1}(-t):=e^{{\frac{i}{\hbar}}tH_{n}},
\end{eqnarray} $\mathcal{U}_{0}(-t)=I$ is a unit operator.
The Hamiltonian $H={\bigoplus\limits}_{n=0}^{\infty}H_{n}$ in \eqref{evol_oper} is a self-adjoint operator with domain
$\mathcal{D}(H)=\{\psi=\oplus_{n=0}^{\infty} \psi_{n}\in{\mathcal{F}_{\mathcal{H}}}\mid \psi_{n}\in\mathcal{D}(H_n)\subset\mathcal{H}_{n},
~{\sum\limits}_{n}\|H_{n}\psi_{n}\|^{2}<\infty\}\subset{\mathcal{F}_{\mathcal{H}}}$ \cite{Kato}.

Assume
$\mathcal{H}=L^{2}(\mathbb{R}^3)$ then an element
$\psi\in\mathcal{F}_{\mathcal{H}}={\bigoplus\limits}_{n=0}^{\infty}L^{2}(\mathbb{R}^{3 n})$ is a sequence of functions
$\psi=\big(\psi_0,\psi_{1}(q_1),\ldots,\psi_{n}(q_1,\ldots,q_{n}),\ldots\big)$
such that $\|\psi\|^{2}=|\psi_0|^{2} + \sum_{n=1}^{\infty}\int dq_1\ldots dq_{n}$ $|\psi_{n}(q_1,\ldots,q_{n})|^{2}<+\infty.$
On the subspace of infinitely
differentiable functions with compact supports $\psi_n\in L^{2}_0(\mathbb{R}^{3 n})\subset L^{2}(\mathbb{R}^{3 n}),$
$n$-particle Hamiltonian $H_{n}$ acts according to the formula ($H_{0}=0$)
\begin{equation}\label{H_Zag}
                H_{n}\psi_n = -\frac{\hbar^{2}}{2}
               \sum\limits_{i=1}^{n}\Delta_{q_i}\psi_n
               +\sum\limits_{k=1}^{n}\sum\limits_{i_{1}<\ldots<i_{k}=1}^{n}\Phi^{(k)}(q_{i_{1}},\ldots,q_{i_{k}})\psi_{n}.
\end{equation}
where  $\Phi^{(k)}$ is a $k$-body interaction potential satisfying Kato conditions
\cite{Kato} and $h={2\pi\hbar}$ is a Planck constant.

We remark that the nature of notations \eqref{evol_oper} used for unitary groups $e^{\pm{\frac{i}{\hbar}}tH_{n}}$
is related to the  correspondence principle between quantum and classical systems \cite{CGP97} and is a consequence
of the existence of two approaches to the description of the evolution of systems in the framework of observables or states.

The properties of a one-parametric family $\{\mathcal{G}(-t)\}_{t\in\mathbb{R}}$ of operators \eqref{groupG} follow from
the properties of groups \eqref{evol_oper} described, for example, in \cite{BerSh}.

\begin{proposition} [\cite{DauL_5},\cite{Pe95}]
On the space $\mathfrak{L}^{1}(\mathcal{F}_\mathcal{H})$
mapping \eqref{groupG}
defines an isometric strongly continuous group, i.e.
one is a $C_{0}$-group, which preserves positivity and self-adjointness of operators.

If $f\in\mathfrak{L}_{0}^{1}(\mathcal{F}_\mathcal{H})\subset\mathcal{D}(-\mathcal{N})\subset\mathfrak{L}^{1}(\mathcal{F}_\mathcal{H})$
then in the sense of the norm convergence of the space $\mathfrak{L}^{1}(\mathcal{F}_\mathcal{H})$ there exists a limit that
is determined the  infinitesimal generator: $-\mathcal{N}=\bigoplus^{\infty}_{n=0}(-\mathcal{N}_{n})$ of group \eqref{groupG}
\begin{equation}\label{infOper}
 \lim\limits_{t\rightarrow 0}\frac{1}{t}\big(\mathcal{G}(-t)f-f\big)=-\frac{i}{\hbar}(Hf-fH):=-\mathcal{N}f,
\end{equation}
 where $H=\bigoplus^{\infty}_{n=0}H_{n}$ is the Hamiltonian \eqref{H_Zag} and
 the operator: $(-i/\hbar)(Hf-fH)$ is defined on the domain $\mathcal{D}(H)\subset\mathcal{F}_\mathcal{H}.$
\end{proposition}

Group of operators \eqref{groupG} and infinitesimal generator \eqref{infOper}
studied within the framework of kernels and symbols of the operators in \cite{BerSh}
and for the Wigner representation in \cite{M}. Some applications of evolution groups
of quantum systems are considered in \cite{AL}.

\subsection{Group of Operators for the Heisenberg Equation}
The adjoint to $\mathfrak{L}^{1}(\mathcal{F}_\mathcal{H})$ space is isometric to
the space $\mathfrak{L}(\mathcal{F}_\mathcal{H})$ of sequences $g=\big(I,g_{1},\ldots,g_{n},\ldots \big)$ of
bounded operators $g_{n}$ ($I$ is a unit operator) defined on the Hilbert space
$\mathcal{H}_n$ satisfying symmetry property
$ g_{n}(1,\ldots,n)=g_{n}(i_1,\ldots,i_n)$ for $\{i_{1},\ldots,i_{n}\}\in \{1,\ldots,n\}$ with an operator norm.
The space $\mathfrak{L}(\mathcal{F}_\mathcal{H})$  is dual to the space $\mathfrak{L}^{1}(\mathcal{F}_\mathcal{H})$
with respect to the bilinear form
\begin{equation}\label{averageD}
        \big\langle g\big|f\big\rangle=
        \sum\limits_{n=0}^{\infty}\frac{1}{n!}
        \text{Tr}_{\mathrm{1,\ldots,n}}~g_{n}f_{n}.
\end{equation}
We will also consider more general space $\mathfrak{L}_{\gamma}(\mathcal{F}_\mathcal{H})$ than $\mathfrak{L}(\mathcal{F}_\mathcal{H})$
with a norm
\begin{eqnarray*}
            \|g\|_{\mathfrak{L}_{\gamma} (\mathcal{F}_\mathcal{H})}=
            \max\limits_{n\geq 0}~ \frac{\gamma^n}{n!}~\|g_{n}\|_{\mathfrak{L}(\mathcal{H}_{n})},
\end{eqnarray*}
where $0<\gamma<1$ and  $\| . \|_{\mathfrak{L}(\mathcal{H}_{n})}$ is an operator norm \cite{Kato}.

An observable of finitely many quantum particles is a sequence of self-adjoint
operators from $\mathfrak{L}_{\gamma}(\mathcal{F}_\mathcal{H})$ and positive normalized
continuous linear functional \eqref{averageD} on the space of observables is interpreted as its mean value \cite{BR}.
The case of the unbounded observables can be reduced to the case under consideration \cite{DauL_5}.

The evolution of observables is described by the initial-value problem
to the Heisenberg equation (the dual von Neumann equation) \cite{DauL_5, Pe95}.
On $\mathfrak{L}(\mathcal{F}_\mathcal{H})$ a solution of the Cauchy problem to the Heisenberg
equation is defined the following one-parametric family of operators \cite{DauL_5}
\begin{eqnarray}\label{grG}
\mathbb{R}^1\ni t\mapsto\mathcal{G}(t)g:=\mathcal{U}(t)g\mathcal{U}^{-1}(t),
\end{eqnarray}
where ~$\mathcal{U}(t)={\bigoplus\limits}_{n=0}^{\infty}\mathcal{U}_{n}(t)$  and the operators
 $\mathcal{U}_{n}(t),~\mathcal{U}_{n}^{-1}(t)$ are defined by formulas \eqref{evol_oper}.
Mapping $\mathcal{G}(t)$ \eqref{grG} is adjoint (dual) to $\mathcal{G}(-t)$ \eqref{groupG}.

\begin{proposition} [\cite{DauL_5},\cite{BR}]
On the space $\mathfrak{L}_{\gamma}(\mathcal{F}_\mathcal{H})$
mapping \eqref{grG} defines an isometric $\ast$-weak continuous group, i.e.
one is a $C_{0}^{\ast}$-group. This group preserves the self-adjointness of operators.
The infinitesimal generator  $\mathcal{N}={\bigoplus\limits}_{n=0}^{\infty}~
\mathcal{N}_{n}$ of this group of operators is a closed operator for the $\ast$-weak topology
and on its domain of the definition $\mathcal{D}(\mathcal{N})\subset\mathfrak{L}_{\gamma}(\mathcal{F}_\mathcal{H})$
which is the everywhere dense set for the $\ast$-weak topology it is
defined in the sense of the $\ast$-weak convergence of the space $\mathfrak{L}_{\gamma}(\mathcal{F}_\mathcal{H})$ by the formula
\begin{equation}\label{infOper1}
 \mathrm{w^{\ast}-}\lim\limits_{t\rightarrow 0}\frac{1}{t}\big(\mathcal{G}(t)g-g\big)=-\frac{i}{\hbar}(gH-Hg),
\end{equation}
where  $H=\bigoplus^{\infty}_{n=0}H_{n}$ is the Hamiltonian \eqref{H_Zag}
and the operator: $\mathcal{N}g=(-i/\hbar)(gH-Hg)$ is defined on the domain $\mathcal{D}(H)\subset\mathcal{F}_\mathcal{H}.$
\end{proposition}

\subsection{Cumulants of Groups of Operators}
Further to the group $\{\mathcal{G}(-t)\}_{t\in\mathbb{R}}$ we will also consider more general mappings
on the space $\mathfrak{L}^{1}_{\alpha}(\mathcal{F}_\mathcal{H})$.
Namely let us  expand  the group $\mathcal{G}(-t)=\bigoplus^{\infty}_{n=0}\mathcal{G}_n(-t)$ as  following
cluster expansions
\begin{eqnarray}\label{groupKlast}
    &\mathcal{G}_{n}(-t,Y)=\sum\limits_{\mathrm{P}:Y ={\bigcup_i} X_i}~
      \prod\limits_{X_i\subset \mathrm{P}}\mathfrak{A}_{|X_i|}(t,X_i),
      \quad\!\!\quad\!\! n = |Y| \geq 0,
\end{eqnarray}
where $ {\sum}_\mathrm{P} $ is the sum over all possible partitions $\mathrm{P}$ of the set $ Y\equiv(1,\ldots,n) $
into $|\mathrm{P}|$ nonempty
mutually disjoint subsets $ X_i\subset Y.$
A solution of  recurrence relations \eqref{groupKlast} is determined by the expansions \cite{GerR}
\begin{eqnarray}\label{cumulant}
        \mathfrak{A}_{n}(t,Y)
        =\sum\limits_{\mathrm{P}:Y ={\bigcup}_i X_i}(-1)^{|\mathrm{P}|-1}(|\mathrm{P}|-1)!
        \prod_{X_i\subset \mathrm{P}}\mathcal{G}_{|X_i|}(-t,X_i),
        \quad\!\!\quad\!\! n = |Y| \geq 0,
\end{eqnarray}
where the notations are similar to that in formula \eqref{groupKlast}.
The operator $\mathfrak{A}_{n}(t)$  we refer to as the $nth$-order cumulant
(semi-invariant) of evolution operators \eqref{groupG}. Some properties of cumulants \eqref{cumulant} considered in \cite{GerSh}.

The generator of the $1st$-order cumulant is given by operator \eqref{infOper}, i.e.
\begin{equation*}
       \lim\limits_{t\rightarrow 0}\frac{1}{t}\big(\mathfrak{A}_{1}(t,Y)-I\big) f_{n}(Y)
       =-\mathcal{N}_{n}(Y)f_{n}(Y),
\end{equation*}
where for $f_n\in\mathfrak{L}_{0}^{1}(\mathcal{H}_n)$ this limit exists
in the sense of the norm convergence of the space $\mathfrak{L}^{1}(\mathcal{H}_n).$

The infinitesimal generator of the $nth$-order cumulant, $n\geq 2,$ is an operator $(-\mathcal{N}^{(n)}_{\mathrm{int}})$
defining by $n$-body interaction potential \eqref{H_Zag}.
According to the equality
\begin{equation}\label{Stirl}
       \sum\limits_{\mathrm{P}:\,Y={\bigcup}_i X_i}
        (-1)^{| \mathrm{P}|-1}(|\mathrm{P}|-1)!=
        \sum\limits_{k=1}^{n}(-1)^{k-1}\mathrm{s}(n,k)(k-1)!=\delta_{n,1},
\end{equation}
where $\mathrm{s}(n,k)$ is the Stirling numbers of the second kind and $\delta_{n,1}$ is a Kroneker symbol,
for the $nth$-order cumulant, $n\geq2,$ in the sense of a point-by-point
convergence of the space $\mathfrak{L}^{1}(\mathcal{H}_{n})$ we have
\begin{eqnarray*}
     \lim\limits_{t\rightarrow 0}\frac{1}{t}\,\mathfrak{A}_{n}(t,Y) f_{n}(Y)
       = \sum\limits_{\mathrm{P}:\,Y={\bigcup}_i X_i}\hskip-2mm
        (-1)^{|\mathrm{P}|-1}(|\mathrm{P}| -1)!\hskip-1mm\sum\limits_{X_i\subset \mathrm{P}}
        (-\mathcal{N}_{|X_i|}(X_i))f_{n}(Y)=\\
        \sum\limits_{\mathrm{P}:\,Y={\bigcup}_i X_i}
        (-1)^{|\mathrm{P}|-1}(|\mathrm{P}| -1)! \sum\limits_{X_i\subset \mathrm{P}}\,\,
        \sum\limits_{k=2}^{\mid X_i\mid}\,\,
        \sum\limits_{i_{1}<\ldots<i_{k}\in\{X_i\}}\big(-\mathcal{N}^{(k)}_{\mathrm{int}}(i_1,\ldots,i_{k})\big)f_{n}(Y).
\end{eqnarray*}
Here for the operator $\Phi^{(n)}$ from Hamiltonian \eqref{H_Zag} the operator $\mathcal{N}^{(n)}_{\mathrm{int}}$ is defined by the formula
\begin{eqnarray}\label{oper Nint2}
      \mathcal{N}^{(n)}_{\mathrm{int}}f_{n}:=-\frac{i}{\hbar}\big(f_{n}\Phi^{(n)}-\Phi^{(n)}f_{n}\big).
\end{eqnarray}
Summing coefficients before every operator $\mathcal{N}^{(k)}_{\mathrm{int}}$ we deduce
\begin{eqnarray}\label{Nint}
        \lim\limits_{t\rightarrow 0}\frac{1}{t}\mathfrak{A}_{n}(t,Y) f_{n}(Y)
       = -\mathcal{N}^{(n)}_{\mathrm{int}}(Y)f_{n}(Y),
\end{eqnarray}
Thus for $f_n\in\mathfrak{L}_{0}^{1}(\mathcal{H}_n)$ the generator of the $nth$-order cumulant is defined by formula \eqref{Nint}
in the sense of the norm convergence of the space $\mathfrak{L}^{1}(\mathcal{H}_n).$

The dual cumulants of the groups $\mathcal{G}(t)=\bigoplus^{\infty}_{n=0}\mathcal{G}_n(t)$ \eqref{grG}
will be introduced in Section 4. For classical many-particle systems cumulants of evolution operators
were introduced in \cite{GerR},\cite{GerRS}.

\section{Group of Operators for the von Neumann Hierarchy}
\setcounter{equation}{0}

The evolution of correlations of quantum finitely many particles is described
by the von Neumann hierarchy for the correlation operators \cite{GerSh}.
It is an equivalent approach for the description of the evolution of all possible states
of quantum many-particle systems in comparison with the approach formulated above.

In what follows we will use such abridged notations: $Y\equiv(1,\ldots,n)$
and $Y_{\mathrm{P}}\equiv(X_{1},\ldots,X_{|\mathrm{P}|})$ is a set
whose elements are $|\mathrm{P}|$ mutually disjoint subsets $X_{i}\subset Y$
of the partition $\mathrm{P}:\,Y={\bigcup\limits}_{i=1}^{|\mathrm{P}|}X_i.$
The $|\mathrm{P}|th$-order cumulant \eqref{cumulant} in this case we denote by $\mathfrak{A}_{|\mathrm{P}|}(t,Y_{\mathrm{P}})$.

A solution of an initial-value problem to the von Neumann hierarchy is defined by a one-parametric family
of nonlinear operators \cite{BM} constructed in \cite{GerSh} with the following properties.

\begin{theorem}
If $f_{n}\in\mathfrak{L}^{1}(\mathcal{H}_{n}),$ $n\geq1,$ then the one-parametric family of nonlinear operators
\begin{equation}\label{groupKum}
      \mathbb{R}^1\ni t\mapsto
       \big(\mathfrak{A}_{t}(f)\big)_{n}(Y):= \sum\limits_{\mathrm{P}:\, Y={\bigcup}_iX_i}
        \mathfrak{A}_{|\mathrm{P}|}(t,Y_{\mathrm{P}})
        \prod_{X_i \subset \mathrm{P}}f_{|{X_i}|}(X_i)
\end{equation}
is a $C_{0}$-group. On the subspace $\mathfrak{L}^{1}_{0}(\mathcal{H}_{n})\subset
    \mathfrak{L}^{1}(\mathcal{H}_{n})$ the infinitesimal generator
    $\mathfrak{N}$ of group \eqref{groupKum} is defined by the operator
\begin{equation}\label{Nnl}
\begin{split}
        &\big(\mathfrak{N}(f)\big)_{n}(Y):=-\mathcal{N}_{n}(Y)f_{n}(Y)+\\
     &\sum\limits_{\mbox{\scriptsize
$\begin{array}{c}\mathrm{P}\hskip-0.7mm: Y={\bigcup}_i X_i,\\|{\mathrm{P}}|>1\end{array}$}}
         \sum\limits_{\mbox{\scriptsize $\begin{array}{c}
       {Z_{1}\subset X_{1}},\\{Z_{1}\neq\emptyset}\end{array}$}}\ldots
        \sum\limits_{\mbox{\scriptsize $\begin{array}{c}
       {Z_{|\mathrm{P}|}\subset X_{|\mathrm{P}|}},\\{Z_{|\mathrm{P}|}\neq\emptyset}\end{array}$}}
        \Big(-\mathcal{N}_{\mathrm{int}}^{\big(\sum\limits_{r=1}^{|\mathrm{P}|}|Z_{{r}}|\big)}
        \big(Z_{{1}},\ldots,Z_{{|\mathrm{P}|}}\big)\Big)\prod_{X_i\subset \mathrm{P}}f_{|{X_i|}}(X_i),
\end{split}
\end{equation}
where the notations are similar to that in \eqref{groupKlast}
and the operator $\mathcal{N}^{(n)}_{\mathrm{int}}$ is defined by formula \eqref{oper Nint2}.
\end{theorem}

\begin{proof} Mapping \eqref{groupKum} is defined for $f_{n}\in\mathfrak{L}^{1}(\mathcal{H}_{n}),$ $n\geq1$,
and the following inequality holds
\begin{eqnarray}\label{es}
        &&\big\|\big(\mathfrak{A}_{t}(f)\big)_{n}\big\|_{\mathfrak{L}^{1}(\mathcal{H}_{n})}\leq n!e^{2n+1}c^{n},
\end{eqnarray}
where $c:=\max\limits_{\mathrm{P}:\,Y={\bigcup}_i X_i}\|f_{|X_{i}|}(X_{i})\|_{_{\mathfrak{L}^{1}(\mathcal{H}_{|X_{i}|})}}.$
Indeed, since for $f_{n}\in \mathfrak{L}^{1}(\mathcal{H}_{n})$  the equality holds \cite{GerSh}
$$\mathrm{Tr}_{1,\ldots,n}|\mathcal{G}_{n}(-t)f_{n}|=\|f_{n}\|_{\mathfrak{L}^{1}(\mathcal{H}_{n})},$$
we have
\begin{equation*}
\begin{split}
        \big\|\big(\mathfrak{A}_{t}(f)\big)_{n}\big\|_{\mathfrak{L}^{1}(\mathcal{H}_{n})}
        \leq \sum\limits_{\mathrm{P}:\,Y={\bigcup}_i X_i}
        \sum\limits_{\mathrm{P}^{'}:\,Y_{\mathrm{P}}={\bigcup}_k Z_k}
 (|\mathrm{P}^{'}|-1)!\prod_{X_i\subset \mathrm{P}}\|f_{|{X_i}|}\|_{\mathfrak{L}^{1}(\mathcal{H}_{|X_{i}|})}\leq\\
     \sum\limits_{\mathrm{P}:\,Y={\bigcup}_i X_i} c^{|\mathrm{P}|}
             \sum\limits_{k=1}^{|\mathrm{P}|}s(|\mathrm{P}|,k)(k-1)!
        \leq \sum\limits_{\mathrm{P}:\,Y={\bigcup}_i X_i} c^{|\mathrm{P}|}
             \sum\limits_{k=1}^{|\mathrm{P}|}k^{|\mathrm{P}|-1}\leq n!e^{2n+1}c^{n},
\end{split}
\end{equation*}
where $\mathrm{s}(|\mathrm{P}|,k)$ are the Stirling numbers of the second kind. That is,  $\big(\mathfrak{A}_{t}(f)\big)_{n}\in \mathfrak{L}^{1}(\mathcal{H}_{n})$ for arbitrary
$t\in\mathbb{R}^{1}$ and  $n\geq 1$.

The group property of a one-parametric family of nonlinear operators \eqref{groupKum}, i.e.
$\mathfrak{A}_{t_{1}}\big(\mathfrak{A}_{t_{2}}(f)\big)=
         \mathfrak{A}_{t_{2}}\big(\mathfrak{A}_{t_{1}}(f)\big)=
         \mathfrak{A}_{t_{1}+t_{2}}(f),$
was proved in \cite{GerSh}.

The strong continuity property of the group $\{\mathfrak{A}_t\}_{t\in\mathbb{R}}$ over the parameter $t\in \mathbb{R}^{1}$
is a consequence of the strong continuity of
group \eqref{groupG} of the von Neumann equation \cite{DauL_5}. Indeed, according to identity \eqref{Stirl}
the following equality holds:
\begin{equation*}
             \sum\limits_{\mathrm{P}:\,Y={\bigcup}_i X_i}\,\,
             \sum\limits_{\mathrm{P}^{'}:\,Y_{\mathrm{P}}={\bigcup}_k Z_k}
        (-1)^{|\mathrm{P}^{'}|-1}(|\mathrm{P}^{'}|-1)!\prod_{X_i\subset \mathrm{P}}f_{|{X_i}|}(X_i)= f_{n}(Y),
\end{equation*}
Therefore, for $f_{n}\in\mathfrak{L}^{1}_{0}(\mathcal{H}_{n}) \subset \mathfrak{L}^{1}(\mathcal{H}_{n})$, $n\geq 1$,
we have
\begin{equation*}
\begin{split}
      &\lim_{t\rightarrow 0}\big\|\big(\mathfrak{A}_{t}(f)\big)_{n}(Y)-f_{n}(Y)
        \big\|_{\mathfrak{L}^{1}(\mathcal{H}_{n})}\leq\\
        &\sum\limits_{\mathrm{P}:\,Y={\bigcup}_i X_i}\,\,
        \sum\limits_{\mathrm{P}^{'}:\,Y_{\mathrm{P}}={\bigcup}_k Z_k}
        \hskip-1mm(|\mathrm{P}^{'}|-1)!\,
         \lim_{t\rightarrow 0}\big\|\big(\hskip-1mm\prod\limits_{Z_{k}\subset\mathrm{P}^{'}}\mathcal{G}_{|Z_{k}|}(-t,Z_{k})-I\big)
         \prod_{X_i\subset \mathrm{P}}f_{|{X_i}|}(X_i)\big\|_{\mathfrak{L}^{1}(\mathcal{H}_{n})}.
\end{split}
\end{equation*}
In view of the the fact that group $\{\mathcal{G}_{n}(-t)\}_{t\in\mathbb{R}}$ \eqref{groupG} is a strong continuous group,
which implies that, for mutually disjoint subsets  $X_{i}\subset Y$, if $f_{n}\in\mathfrak{L}_{0}^{1}(\mathcal{H}_{n})\subset\mathfrak{L}^{1}(\mathcal{H}_{n})$ in the sense of the norm convergence
$\mathfrak{L}^{1}(\mathcal{H}_{n})$ there exists the limit
\begin{eqnarray*}
\lim_{t\rightarrow 0}\big(\prod\limits_{Z_{k}\subset\mathrm{P}^{'}}\mathcal{G}_{|Z_{k}|}(-t,Z_{k})f_{n}
-f_{n}\big)=0.
\end{eqnarray*}
Thus if $f\in\mathfrak{L}^{1}(\mathcal{F}_\mathcal{H})$ we finally obtain
\begin{equation*}
\lim_{t\rightarrow 0}\big\|\big(\mathfrak{A}_{t}(f)\big)_{n}-f_{n}\big\|_{\mathfrak{L}^{1}(\mathcal{H}_{n})}=0.
\end{equation*}

We now construct the infinitesimal generator $\mathfrak{N}$ of group \eqref{groupKum}. Taking into account that for
$f_{n}\in\mathfrak{L}_{0}^{1}(\mathcal{H}_{n})$ equality \eqref{infOper} holds,
we differentiate the $|\mathrm{P}|th$-order cumulant $\mathfrak{A}_{|\mathrm{P}|}(t,Y_{\mathrm{P}})$
for all  $\psi_{n}\in \mathcal{D}(H_{n})\subset\mathcal{H}_{n}$ in the sense of the point-by-point convergence.
According to equality \eqref{Nint} for  $|\mathrm{P}|\geq 2$ we derive
\begin{equation}\label{derivation}
\begin{split}
       &\lim\limits_{t\rightarrow 0}\frac{1}{t}\mathfrak{A}_{|\mathrm{P}|}(t,Y_{\mathrm{P}})f_{n} \psi_{n}
       =\sum\limits_{\mathrm{P}^{'}:\,Y_{\mathrm{P}}={\bigcup}_k Z_k}
        (-1)^{|\mathrm{P}^{'}|-1}(|\mathrm{P}^{'}| -1)!\sum\limits_{Z_k\subset \mathrm{P}^{'}}
        (-\mathcal{N}_{|Z_{k}|}(Z_k))f_{n}\psi_{n}\\
        &=\sum\limits_{\mbox{\scriptsize $\begin{array}{c}
       {Z_{1}\subset X_{1}},\\{Z_{1}\neq\emptyset}\end{array}$}}\ldots
        \sum\limits_{\mbox{\scriptsize $\begin{array}{c}
       {Z_{|\mathrm{P}|}\subset X_{|\mathrm{P}|}},\\{Z_{|\mathrm{P}|}\neq\emptyset}\end{array}$}}
        \Big(-\mathcal{N}_{\mathrm{int}}^{\big(\sum\limits_{r=1}^{|\mathrm{P}|}|Z_{{r}}|\big)}
        \big(Z_{{1}},\ldots,Z_{{|\mathrm{P}|}}\big)\Big)f_{n}\psi_{n},
\end{split}
\end{equation}
where ${\sum\limits}_{\substack{Z_{j}\subset X_{j}}}$  is a sum over all subsets
$Z_{j}\subset X_{j}$ of the set $ X_{j}$.
Then in view of equality \eqref{derivation} for group \eqref{groupKum} we obtain
\begin{equation*}
\begin{split}
     &\lim\limits_{t\rightarrow 0}\frac{1}{t}\Big(\big(\mathfrak{A}_{t}(f)\big)_{n}-f_{n}\Big)\psi_{n}
     =\lim\limits_{t\rightarrow 0}\frac{1}{t}\Big(\sum\limits_{\mathrm{P}:\,Y={\bigcup}_i X_i}
                \mathfrak{A}_{|\mathrm{P}|}(t,Y_{\mathrm{P}})
                \prod_{X_i\subset \mathrm{P}}f_{|{X_i}|}(X_i)-f_{n}(Y)\Big)\psi_{n}\\
         &=\lim\limits_{t\rightarrow 0}\frac{1}{t}\big(\mathfrak{A}_{1}(t,Y)f_{n}-f_{n}\big)\psi_{n}
            +  \sum\limits_{\mbox{\scriptsize $\begin{array}{c}
       {\mathrm{P}}:Y={\bigcup}_i X_i,\\|{\mathrm{P}}|>1
       \end{array}$}}\lim\limits_{t\rightarrow 0}\frac{1}{t}\mathfrak{A}_{|\mathrm{P}|}(t,Y_{\mathrm{P}})
       \prod_{X_i\subset \mathrm{P}}f_{|{X_i}|}(X_i)\psi_{n}\\
         &=\big(-\mathcal{N}_{n}f_{n}\big)(Y)\psi_{n} \\
         &+\sum\limits_{\mbox{\scriptsize $\begin{array}{c}
       \mathrm{P}:Y={\bigcup}_i X_i,\\|{\mathrm{P}}|>1 \end{array}$}}
       \sum\limits_{\mbox{\scriptsize $\begin{array}{c}
       {Z_{1}\subset X_{1}},\\{Z_{1}\neq\emptyset}\end{array}$}}\ldots
        \sum\limits_{\mbox{\scriptsize $\begin{array}{c}
       {Z_{|\mathrm{P}|}\subset X_{|\mathrm{P}|}},\\{Z_{|\mathrm{P}|}\neq\emptyset}\end{array}$}}
        \Big(-\mathcal{N}_{\mathrm{int}}^{\big(\sum\limits_{r=1}^{|\mathrm{P}|}|Z_{{r}}|\big)}
        \big(Z_{{1}},\ldots,Z_{{|\mathrm{P}|}}\big)\Big)\prod_{X_i\subset \mathrm{P}}f_{|{X_i}|}(X_i)\psi_{n}.
\end{split}
\end{equation*}
Thus for $f\in\mathfrak{L}_{0}^{1}(\mathcal{F}_\mathcal{H})\subset\mathcal{D}
 (\mathfrak{N})\subset\mathfrak{L}^{1}(\mathcal{F}_\mathcal{H})$ in the sense of the norm convergence
 $\mathfrak{L}^{1}(\mathcal{H}_{n})$
  we finally have
$$ \lim_{t\rightarrow 0}\big\| \frac{1}{t}\big(\big(\mathfrak{A}_{t}(f)\big)_{n}-
 f_{n}\big)
 -\big(\mathfrak{N}(f)\big)_{n}\big\|_{\mathfrak{L}^{1}(\mathcal{H}_{n})}=0.\eqno\qedhere$$
\end{proof}

We give an example which illustrates the structure of expansion \eqref{groupKum}.
For the correlation operators $f=\big(0,f_{1}(1),0,\ldots\big)$ that is interpreted as satisfying the "chaos" property \cite{CGP97},
we have
\begin{equation*}
       \big(\mathfrak{A}_{t}(f)\big)_{n}= \mathfrak{A}_n(t,1,\ldots,n)
        \prod_{i=1}^n f_{1}(i),\quad n\geq1,
\end{equation*}
i.e. if at the initial instant there are no correlations in a system,
the correlations generated by the dynamics of a system are completely governed by cumulants
of groups \eqref{groupG}.

We now consider the structure of infinitesimal generator \eqref{Nnl} for a two-body interaction potential
\begin{equation*}
        \big(\mathfrak{N}(f)\big)_{n}(Y)=-\mathcal{N}_{n}(Y)f_{n}(Y)+
     \sum\limits_{\mathrm{P}:\,Y=X_1\bigcup X_2}~
      \sum\limits_{i_1\in\{X_1\}}\sum\limits_{i_2\in\{X_2\}} \big(-\mathcal{N}_{\mathrm{int}}^{(2)}(i_1,i_2)\big)
f_{|{X_1|}}(X_1)f_{|{X_2|}}(X_2)\nonumber,
\end{equation*}
where the symbol ${\sum\limits}_{\mathrm{P}:\,Y=X_1\bigcup X_2}$ means summation over all partitions of the set $Y$
into two nonempty parts $X_1$ and $X_2$
and the operator $\mathcal{N}^{(2)}_{\mathrm{int}}$ is defined by formula \eqref{oper Nint2}.
For classical systems this generator is an equivalent notation of the generator of the
Liouville hierarchy \cite{Sh} formulated in \cite{Gre56}.

\section{Group of Operators for the Quantum BBGKY Hierarchy}
\setcounter{equation}{0}

The evolution of all possible states both finitely and infinitely many quantum particles is described by the initial-value problem
to the BBGKY hierarchy for marginal density operators \cite{Pe95},\cite{GerS}.
For finitely many particles this hierarchy of equations is an equivalent to the von Neumann equation.

We will use notations from the previous section. Since $Y_{\mathrm{P}}\equiv(X_1,\ldots,X_{|\mathrm{P}|})$
then $Y_{1}$ is the set consisting of one element of the partition $\mathrm{P}$ ($|\mathrm{P}|=1$) of the set $Y\equiv(1,\ldots,s)$.
In this case for $n \geq 0$ the $(1+n)th$-order cumulant  of operators \eqref{groupG} is defined by the formula
\begin{eqnarray}\label{cum}
\mathfrak{A}_{1+n}(t,Y_{1},X \backslash Y):=
\sum\limits_{\mathrm{P}:\,\{Y_{1},X\setminus Y\}=
{\bigcup}_i X_i}(-1)^{|\mathrm{P}|-1}(|\mathrm{P}|-1)!
        \prod_{X_i\subset \mathrm{P}}\mathcal{G}_{|X_i|}(-t,X_i),
\end{eqnarray}
where ${\sum}_\mathrm{P} $
    is the sum over all possible partitions $\mathrm{P}$ of the set $\{Y_{1},X\setminus Y\}=\{Y_{1},s+1,\ldots,s+n\}$ into
    $|\mathrm{P}|$ nonempty mutually disjoint subsets
    $ X_i\subset \{Y_{1},X\setminus Y\}$.

On the space $\mathfrak{L}_{\alpha}^{1}(\mathcal{F} _\mathcal{H})$ a solution of the
initial-value problem to the BBGKY hierarchy is defined by a one-parametric mapping \cite{GerS} with the following properties.

\begin{theorem}
If $f\in\mathfrak{L}_{\alpha}^{1}(\mathcal{F} _\mathcal{H})$ and $\alpha>e$, then the one-parametric mapping
\begin{eqnarray}\label{RozvBBGKY}
 \mathbb{R}^1\ni t\mapsto(U(t)f) _{s}(Y):=
  \sum\limits_{n=0}^{\infty}\frac{1}{n!} \mathrm{Tr}_{\mathrm{s+1,\ldots,{s+n}}}
      \mathfrak{A}_{1+n}(t,Y_{1},X \backslash Y)f_{s+n}(X)
\end{eqnarray}
is a $C_{0}$-group.
On the subspace $\mathfrak{L}^{1}_{\alpha, 0}\subset\mathfrak{L}^{1}_{\alpha}(\mathcal{F}_\mathcal{H})$ the infinitesimal generator
${\mathfrak{B}}={\bigoplus\limits}_{n=0}^{\infty}
{\mathfrak{B}}_{n}$ of group \eqref{RozvBBGKY} is defined by the operator $(s\geq 1)$
\begin{equation} \label{1}
\begin{split}
        &(\mathfrak{B}f)_{s}(Y):=\\
        &-\mathcal{N}_{s}(Y)f_{s}(Y)+
        \sum\limits_{k=1}^{s}\frac{1}{k!}\sum\limits_{i_1\neq\ldots\neq i_{k}=1}^{s}
        \,\sum\limits_{n=1}^{\infty}\frac{1}{n!}\mathrm{Tr}_{\mathrm{s+1,\ldots,s+n}}
    \big(-\mathcal{N}_{\mathrm{int}}^{(k+n)}\big)(i_1,\ldots, i_{k},X \backslash Y)f_{s+n}(X),
\end{split}
\end{equation}
where on $\mathfrak{L}_{0}^{1}(\mathcal{H}_{s+n})\subset\mathfrak{L}^{1}(\mathcal{H}_{s+n})$
the operator $\mathcal{N}^{(k+n)}_{\mathrm{int}}$ is defined by formula \eqref{oper Nint2}.
\end{theorem}

\begin{proof}
If $f\in\mathfrak{L}_{\alpha}^{1}(\mathcal{F} _\mathcal{H})$ mapping \eqref{RozvBBGKY} is defined
provided that $\alpha> e $ and the following estimate holds \cite{GerS}
\begin{eqnarray*}\label{estim_L_{1}}
        \|U(t)f\|_{\mathfrak{L}_{\alpha}^{1}(\mathcal{F}_\mathcal{H})}\leq
        c_{\alpha}\|f \|_{\mathfrak{L}_{\alpha}^{1}(\mathcal{F}_\mathcal{H})},
\end{eqnarray*}
where $c_{\alpha}=e^{2}(1-\frac{e}{\alpha})^{-1}$ is a constant. Similar to \eqref{es}
this estimate comes out from the inequality for cumulant \eqref{cum}
\begin{eqnarray*}\label{estim_L_{s+n}}
        \|\mathfrak{A}_{1+n}(t)f_{s+n}\|_{\mathfrak{L}^{1}(\mathcal{H}_{s+n})}\leq
        n!e^{n+2}\|f_{s+n}\|_{\mathfrak{L}^{1}(\mathcal{H}_{s+n})}.
\end{eqnarray*}

The strong continuity property of the group $U(t)$ over the parameter $t\in \mathbb{R}^{1}$
is a consequence of the strong continuity of
group \eqref{groupG} of the von Neumann equation.

We now construct an infinitesimal generator of group \eqref{RozvBBGKY}. Taking into account that for
$f_{n}\in\mathfrak{L}_{0}^{1}(\mathcal{H}_{n})$ equality \eqref{infOper} holds,
we differentiate the expression of cumulant \eqref{cum} in the sense of the point-by-point convergence.
According to equality \eqref{derivation} for $n\geq 1$, we derive
\begin{equation}\label{derivation2}
\begin{split}
       &\lim\limits_{t\rightarrow 0}\frac{1}{t}\mathfrak{A}_{1+n}(t,Y_{1},X \backslash Y)f_{s+n} \psi_{s+n}=
       \sum\limits_{\mbox{\scriptsize $\begin{array}{c}{Z \subset Y},\\{Z \neq\emptyset}\end{array}$}}
       \big(-\mathcal{N}_{\mathrm{int}}^{(|Z|+n)}\big)(Z,s+1,\ldots,s+n)
        f_{s+n}\psi_{s+n}=\\
        &\sum\limits_{k=1}^{|Y|}\frac{1}{k!}\sum\limits_{i_1\neq\ldots\neq i_{k}=1}^{|Y|}
        \big(-\mathcal{N}_{\mathrm{int}}^{(k+n)}\big)(i_1,\ldots,i_{k},X \backslash Y)f_{s+n}\psi_{s+n},
\end{split}
\end{equation}
where ${\sum\limits}_{\substack{Z\subset X}}$ is a sum over all subsets $Z\subset X$ of the set $X$.
Then taking into account formula \eqref{infOper} for $n=1$ and  equality  \eqref{derivation2} for group \eqref{RozvBBGKY} we obtain
\begin{equation*}
\begin{split}
        &\lim\limits_{t\rightarrow 0}\frac{1}{t}\big(\big(U(t)f\big)_{s}-f_{s}\big)\psi_{s}=\\
        &\lim\limits_{t\rightarrow 0}\frac{1}{t}\big(\mathfrak{A}_{1}(t,Y)f_{s}-f_{s}\big)\psi_{s}
        +  \sum\limits_{n=1}^{\infty}\frac{1}{n!}\mathrm{Tr}_{\mathrm{s+1,\ldots,{s+n}}}
        \lim\limits_{t\rightarrow 0}\frac{1}{t}\mathfrak{A}_{1+n}(t,Y_{1},X \backslash Y)f_{s+n}\psi_{s}=\\
        &-\mathcal{N}_{s}f_{s}\psi_{s}+
        \sum\limits_{n=1}^{\infty}\frac{1}{n!}\mathrm{Tr}_{\mathrm{s+1,\ldots,{s+n}}}
        \sum\limits_{k=1}^{s}\frac{1}{k!}\sum\limits_{i_1\neq\ldots\neq i_{k}=1}^{s}
        \big(-\mathcal{N}_{\mathrm{int}}^{(k+n)}\big)(i_1,\ldots,i_{k},X \backslash Y)\psi_{s}.
\end{split}
\end{equation*}
Thus, if $\mathfrak{L}^{1}_{\alpha,0}\subset\mathcal{D}
 (\mathfrak{B})\subset\mathfrak{L}^{1}_{\alpha}(\mathcal{F}_\mathcal{H})$
we finally have in the sense of the norm convergence
\begin{eqnarray*}
   \lim_{t\rightarrow 0}\big\| \frac{1}{t}\big(U(t)f- f\big)
   -\mathfrak{B}f\big\|_{\mathfrak{L}^{1}_{\alpha}(\mathcal{F}_\mathcal{H})}=0,
\end{eqnarray*}
where the operator $\mathfrak{B}$ on $\mathfrak{L}^{1}_{\alpha,0}$ is given by formula \eqref{1}.
\end{proof}

We now give an example which illustrates the structure of infinitesimal generator \eqref{1}. In the case of
two-body interaction potential \eqref{H_Zag} operator \eqref{1} has the form ($s\geq 1$)
\begin{eqnarray}\label{bg}
        (\mathfrak{B}f)_{s}(Y) = -\mathcal{N}_{s}(Y)f_{s}(Y)+
        \sum\limits_{i=1}^{s} \mathrm{Tr}_{\mathrm{s+1}}
        \big(-\mathcal{N}_{\mathrm{int}}^{(2)}\big)(i,s+1)f_{s+1}(Y,s+1),
\end{eqnarray}
where the operator $\mathcal{N}^{(2)}_{int}$ is defined
on $\mathfrak{L}_{0}^{1}(\mathcal{H}_{s+1})\subset\mathfrak{L}^{1}(\mathcal{H}_{s+1})$
by formula \eqref{oper Nint2} for $n=2$.
For $\mathcal{H}=L^{2}(\mathbb{R}^3)$ in the framework
of kernels of operators $f_{s}$ ($s$-particle density matrix or marginal distributions \cite{Pe95}) operator \eqref{bg} takes
a canonical form of a generator of the quantum BBGKY hierarchy \cite{CGP97},\cite{GerS}
\begin{equation*}
\begin{split}
      &(\mathfrak{B}f)_{s}(q_{1},\ldots,q_{s};q^{'}_{1},\ldots,q^{'}_{s})= \\
      &-\frac{i}{\hbar}\Big(-\frac{\hbar^{2}}{2}\sum\limits_{i=1}^{s}(\Delta_{q_{i}}-\Delta_{q^{'}_{i}})+
      \sum\limits_{i<j=1}^{s}\big(\Phi^{(2)}(q_{i}-q_{j})-\Phi^{(2)}(q^{'}_{i}-q^{'}_{j})\big)\Big)
      f_{s}(q_{1},\ldots,q_{s};q^{'}_{1},\ldots,q^{'}_{s})-\\
      &\frac{i}{\hbar}\sum\limits_{i=1}^{s}\int dq_{s+1}\big(
      \Phi^{(2)}(q_{i}-q_{s+1})-\Phi^{(2)}(q^{'}_{i}-q_{s+1})\big)f_{s+1}(q_{1},\ldots
      q_{s},q_{s+1};q^{'}_{1},\ldots,q^{'}_{s},q_{s+1}).
\end{split}
\end{equation*}

In \cite{Pe95},\cite{GerS} for quantum and in the book \cite{CGP97} for classical systems of particles
with a two-body interaction potential, an equivalent representation for group \eqref{RozvBBGKY} was used,
namely for the case under consideration the group $U(t)$ has the following representation
\begin{eqnarray}\label{rgb}
U(t)=\mathcal{G}(-t)+ \sum\limits_{n=1}^{\infty}\frac{1}{n!}
\big[\underbrace{\mathfrak{a},\ldots,\big[{\mathfrak{a}}}_{\hbox{n-times}},\mathcal{G}(-t)\big]\ldots \big]=
 e^{\mathfrak{a}}\mathcal{G}(-t)e^{-\mathfrak{a}},
\end{eqnarray}
where $\big[\, .\, , \,.\,\big]$ is a commutator, the operator $\mathfrak{a}$ (an analog of the annihilation
operator) is defined on the space $\mathfrak{L}_{\alpha}^{1}(\mathcal{F}_\mathcal{H})$ by the formula
\begin{eqnarray}\label{oper_an}
         \big( \mathfrak{a} f \big)_{s}(Y):=\mathrm{Tr}_{\mathrm{s+1}} f_{s+1}(Y,s+1),
\end{eqnarray}
the operators $e^{\pm \mathfrak{a}}$ are defined on the space $\mathfrak{L}_{\alpha}^{1}(\mathcal{F}_\mathcal{H})$
by the expansions
\begin{eqnarray*}
\big(e^{\pm \mathfrak{a}}f\big)_s(Y)=\sum_{n=0}^{\infty}\,\frac{(\pm 1)^n}{n!}\mathrm{Tr}_{\mathrm{s+1,\ldots,s+n}}f_{s+n}(Y,s+1,\ldots,s+n) ,
           \quad s\geq 1.
\end{eqnarray*}
and the group of operators $\mathcal{G}(-t)$ is defined by expression \eqref{groupG}.
Representation \eqref{rgb} is true in consequence of definition \eqref{oper_an} of the operator $\mathfrak{a}$
and the validity of an equality for every $s\geq 1$
\begin{eqnarray*}
\big(\big[\underbrace{\mathfrak{a},\ldots,\big[{\mathfrak{a}}}_{\hbox{n-times}},\mathcal{G}(-t)\big]\ldots \big]f\big)_s(Y)=
\mathrm{Tr}_{\mathrm{s+1,\ldots,{s+n}}}
      \mathfrak{A}_{1+n}(t,Y_{1},X \backslash Y)f_{s+n}(X),
\end{eqnarray*}
where the notations are similar to that in \eqref{cum}.
For example, since for an isometric group $\mathcal{G}(-t)$ and trace class operators $f$
the following equality holds
\begin{eqnarray}\label{transf}
\mathrm{Tr}_{\mathrm{s+1}}\mathcal{G}_{1}(-t,s+1)f_{s+1}(Y,s+1)= \mathrm{Tr}_{\mathrm{s+1}}f_{s+1}(Y,s+1).
\end{eqnarray}
then if $n=1$ we have
\begin{equation*}
\begin{split}
&\big(\big[\mathfrak{a},\mathcal{G}(-t)\big]f\big)_s(Y)=
\mathrm{Tr}_{\mathrm{s+1}}\big(\mathcal{G}_{s+1}(-t,Y,s+1)f_{s+1}(Y,s+1)-\mathcal{G}_{s}(-t,Y)f_{s+1}(Y,s+1)\big)=\\
&\mathrm{Tr}_{\mathrm{s+1}}\mathfrak{A}_{2}(t,Y_{1},s+1)f_{s+1}(Y,s+1).
\end{split}
\end{equation*}

We note that cluster expansions \eqref{groupKlast} can be put into basis
of all possible representations of a group of operators for the quantum BBGKY hierarchy. In fact
representation \eqref{rgb} we obtain solving
recurrence relations \eqref{groupKlast} with respect to the $1st$-order
cumulants for the separation terms which are independent from the variable $Y$
\begin{eqnarray*}
   \mathfrak{A}_{1+n}\big(t,Y_1,X\backslash Y\big)=
    \sum\limits_{Z\subset X\backslash Y}\mathfrak{A}_{1}(t,Y\cup
       Z)\sum\limits_{\mathrm{P}:\,(X\backslash Y)\backslash Z ={\bigcup\limits}_i X_i}\,
       (-1)^{|\mathrm{P}|}\,|\mathrm{P}|!\,\,
    {\prod\limits}_{i=1}^{|\mathrm{P}|}\mathfrak{A}_{1}(t,X_{i}),
\end{eqnarray*}
where ${\sum\limits}_{\substack{Z\subset X\backslash Y}}$ is a sum over all subsets $Z\subset X\backslash Y$ of the set $X\backslash Y$.
Then taking into account equality \eqref{transf} and
\begin{eqnarray}\label{eq1}
 \sum\limits_{\mathrm{P}:\,(X\backslash Y)\backslash Z =
 {\bigcup\limits}_i X_i}(-1)^{|\mathrm{P}|}\,|\mathrm{P}|!=(-1)^{|(X\backslash Y)\backslash Z|},
\end{eqnarray}
for the group $U(t)$ we have
\begin{eqnarray*}\label{cherez1}
    \big(U(t)f\big)_s(Y)=\sum\limits_{n=0}^{\infty}\frac{1}{n!}
    \mathrm{Tr}_{s+1,\ldots,s+n}\sum\limits_{Z\subset X\backslash
    Y}(-1)^{|(X\backslash Y)\backslash Z|}\mathcal{G}_{|Y\cup Z|}(-t,Y\cup Z)f_{|X|}(X).
    \end{eqnarray*}
Thus as a result of the symmetry property and definition \eqref{oper_an} of the operator $\mathfrak{a}$ we derive \eqref{rgb}
\begin{eqnarray*}\label{Pf}
       U(t)=\sum\limits_{n=0}^{\infty}\frac{1}{n!}\sum\limits_{k=0}^{n}(-1)^{k}\frac{n!}{k!(n-k)!}
       \mathfrak{a}^{n-k}\mathcal{G}(-t)\mathfrak{a}^{k}=e^{\mathfrak{a}}\mathcal{G}(-t)e^{-\mathfrak{a}}.
\end{eqnarray*}

We can obtain one more representation of the group $U(t)$ solving
recurrence relations \eqref{groupKlast}  with respect to the $1st$-order and $2nd$-order
cumulants. In fact if $n\geq1$ we get
\begin{eqnarray*}
    \mathfrak{A}_{1+n}(t,Y_1,X\backslash Y)=
    \sum\limits_{\substack{Z\subset{X\backslash
    Y,}\\Z\neq \emptyset}}\mathfrak{A}_{2}(t,Y,Z)
    \sum\limits_{\mathrm{P}:\,(X\backslash Y)\backslash Z ={\bigcup\limits}_i X_i}
    (-1)^{|\mathrm{P}|}\,|\mathrm{P}|!\,\, {\prod}_{i=1}^{|\mathrm{P}|}\mathfrak{A}_{1}(t,X_{i}).
\end{eqnarray*}
Then taking into account equalities \eqref{transf} and \eqref{eq1} we derive
\begin{eqnarray*}
    &&\big(U(t)f\big)_s(Y)=\mathfrak{A}_{1}(t,Y)f_{s}(Y)+\sum\limits_{n=1}^{\infty}\frac{1}{n!}\mathrm{Tr}_{\mathrm{s+1,\ldots,s+n}}
    \sum\limits_{\substack{Z\subset{X\backslash
    Y,}\\Z\neq \emptyset}}(-1)^{|(X\backslash Y)\backslash Z|}
    \mathfrak{A}_{2}(t,Y,Z)f_{s+n}(X).
\end{eqnarray*}

An infinitesimal generator of a group of operators \eqref{rgb} has the following form
\begin{eqnarray}\label{rbg}
{\mathfrak{B}}=-\mathcal{N}+ \sum\limits_{n=1}^{\infty}\frac{1}{n!}
\big[\underbrace{\mathfrak{a},\ldots,\big[{\mathfrak{a}}}_{\hbox{n-times}},(-\mathcal{N})\big]\ldots \big]=
e^{\mathfrak{a}}(-\mathcal{N})e^{-\mathfrak{a}}.
\end{eqnarray}
Representation \eqref{rbg} is true in consequence of definition \eqref{oper_an} of the operator $\mathfrak{a}$
and the validity of equalities
\begin{equation*}
\begin{split}
&\big(\big[\underbrace{\mathfrak{a},\ldots,\big[{\mathfrak{a}}}_{\hbox{n-times}},(-\mathcal{N})\big]\ldots \big]f\big)_s(Y)=
\sum\limits_{k=0}^{n}(-1)^{k}\frac{n!}{k!(n-k)!}\big(\mathfrak{a}^{n-k}\big(\mathcal{-N}\big)\mathfrak{a}^{k}f\big)_{s}(Y)=\\
&\sum\limits_{k=0}^{n}(-1)^{k}\sum\limits_{i_{1}<\ldots<i_{n-k}=s+1}^{s+n}
 \mathrm{Tr}_{\mathrm{s+1,\ldots,s+n}} \big(\mathcal{-N}_{s+n-k}(Y,i_{1},\ldots,i_{n-k})\big)f_{s+n}(X)=\\
&\sum\limits_{k=1}^{s}\frac{1}{k!}\sum\limits_{i_1\neq\ldots\neq i_{k}=1}^{s}
\mathrm{Tr}_{\mathrm{s+1,\ldots,s+n}}\big(-\mathcal{N}_{\mathrm{int}}^{(k+n)}\big)(i_1,\ldots, i_{k},X \backslash Y)f_{s+n}(X).
\end{split}
\end{equation*}

For a system of particles interacting through a two-body potential this equality reduces to the following one
\begin{eqnarray*}
\big(\big[\mathcal{N},\mathfrak{a}\big]f\big)_s(Y)=
\sum\limits_{i=1}^{s} \mathrm{Tr}_{\mathrm{s+1}}
        \big(-\mathcal{N}_{\mathrm{int}}^{(2)}\big)(i,s+1)f_{s+1}(Y,s+1),
\end{eqnarray*}
that is true in view of the equality: $\mathrm{Tr}_{\mathrm{s+1}}\mathcal{N}_{1}(s+1)f_{s+1}(Y,s+1)=0$.

\section{Group of Operators for the Quantum Dual BBGKY Hierarchy}
\setcounter{equation}{0}

The evolution of marginal observables of both finitely and infinitely many quantum  particles
is described by the initial-value problem
to the dual BBGKY hierarchy. This hierarchy of equations is dual to the quantum BBGKY hierarchy
in the sense of bilinear form \eqref{averageD} and for finitely many particles one is an equivalent
to the Heisenberg equation (the dual von Neumann equation). For systems of classical particles
the dual BBGKY hierarchy was examined in \cite{CGP97},\cite{BG},\cite{GerR}.

In this section we will use such abridged notations: $Y\equiv(1,\ldots,s)$, $X\equiv Y\backslash\{j_1,\ldots,j_{s-n}\}$.
According to notations of section 2, the set $(Y\backslash X)_1$ consists of one element $Y\backslash X=(j_1,\ldots,j_{s-n})$,
i.e. the set $(j_1,\ldots,j_{s-n})$ is connected subset of the partition $\mathrm{P}$ ($|\mathrm{P}|=1$).
In the case under consideration the dual cumulants $\mathfrak{A}_{1+n}^+(t), ~n\geq0,$ of groups \eqref{grG} are defined by the formula
\begin{equation} \label{rozv_rec}
    \mathfrak{A}^+_{1+n}\big(t,(Y\backslash X)_1, X\big):=
    \sum\limits_{\mathrm{P}:\,\{(Y\backslash X)_1, X\}={\bigcup}_i X_i}
    (-1)^{\mathrm{|P|}-1}({\mathrm{|P|}-1})!\prod_{X_i\subset \mathrm{P}}\mathcal{G}_{|X_i|}(t,X_i),
\end{equation}
where ${\sum}_\mathrm{P} $
is the sum over all possible partitions $\mathrm{P}$ of the set $\{(Y\backslash X)_1,j_1,\ldots,j_{s-n}\}$ into
$|\mathrm{P}|$ nonempty mutually disjoint subsets  $ X_i\subset \{(Y\backslash X)_1, X\}$.

On the space $\mathfrak{L}_{\gamma}(\mathcal{F}_\mathcal{H})$ a solution of the initial-value
problem to the dual BBGKY hierarchy is defined by a one-parametric mapping (the adjoint mapping to \eqref{RozvBBGKY}
in the sense of bilinear form \eqref{averageD}) with the following properties.

\begin{theorem}
If $g\in\mathfrak{L}_{\gamma}(\mathcal{F}_\mathcal{H})$ and $\gamma<e^{-1}$, then the one-parametric mapping
\begin{equation}\label{krut_rozv}
\begin{split}
       &\mathbb{R}^{1}\ni t\mapsto\big(U^{+}(t)g\big) _{s}(Y):=\\
      &\sum_{n=0}^s\,\frac{1}{(s-n)!}\sum_{j_1\neq\ldots\neq j_{s-n}=1}^s
      \mathfrak{A}_{1+n}^+ \big(t,(Y\backslash X)_1,X\big) \,
      g_{s-n}(Y\backslash X),  \quad\!\! s\geq 1
\end{split}
\end{equation}
is a $C_{0}^{\ast}$-group.
The infinitesimal generator  ${\mathfrak{B}}^{+}={\bigoplus\limits}_{n=0}^{\infty}
{\mathfrak{B}}^{+}_{n}$  of this group of operators is a closed operator for the $\ast$-weak topology and
on the domain of the definition $\mathcal{D}({\mathfrak{B}}^{+})\subset\mathfrak{L}_{\gamma}(\mathcal{F}_\mathcal{H})$
which is the everywhere dense set for the $\ast$-weak topology of the space $\mathfrak{L}_{\gamma}(\mathcal{F}_\mathcal{H})$
it is defined by the operator
\begin{eqnarray}\label{d}
\begin{split}
       &({\mathfrak{B}}^{+} g)_{s}(Y):= \mathcal{N}_{s}(Y)g_{s}(Y)+\\
       &\sum\limits_{n=1}^{s}\frac{1}{n!}
        \sum\limits_{k=n+1}^s \frac{1}{(k-n)!}\sum_{j_1\neq\ldots\neq j_{k}=1}^s
        \mathcal{N}_{\mathrm{int}}^{(k)}(j_1,\ldots,j_{k})
        g_{s-n}(Y\backslash\{j_1,\ldots,j_{n}\}),
\end{split}
\end{eqnarray}
where the operator $\mathcal{N}^{(k)}_{\mathrm{int}}$ is given by formula \eqref{oper Nint2}.
\end{theorem}

\begin{proof}
If $g\in\mathfrak{L}_{\gamma}(\mathcal{F}_\mathcal{H})$ mapping \eqref{krut_rozv} is defined
provided that $\gamma<e^{-1}$ and the following estimate holds
\begin{eqnarray*}
  \big\|U^{+}(t)g\big\|_{\mathfrak{L}_{\gamma}(\mathcal{F}_\mathcal{H})}
  \leq e^2(1-\gamma e)^{-1}\|g\|_{\mathfrak{L}_{\gamma}(\mathcal{F}_\mathcal{H})}.
\end{eqnarray*}
Similar to \eqref{es} this estimate comes out from the inequality
\begin{equation*}
\begin{split}
  &\big\|(U^{+}(t)g)_s\big\|_{\mathfrak{L}(\mathcal{H}_s)}\leq
  \sum_{n=0}^s\,\frac{1}{(s-n)!}\sum_{j_1\neq\ldots\neq j_{s-n}=1}^s\,
  \sum_{\mathrm{P}:\{Y\backslash X)_1,X\} =\bigcup_i X_i}(|\mathrm{P}|-1)!
  \|g_{s-n}\|_{\mathfrak{L}(\mathcal{H}_{s-n})}\leq\\
  &\sum_{n=0}^s \|g_{s-n}\|_{\mathfrak{L}(\mathcal{H}_{s-n})}\frac{s!}{n!(s-n)!}
  \sum_{k=1}^{n+1}s(n+1,k)(k-1)!,
\end{split}
\end{equation*}
where $s(n+1,k)$ is the Stirling numbers of the second kind.

On the space $\mathfrak{L}_{\gamma}(\mathcal{F}_\mathcal{H})$
the $\ast$-weak continuity property  of the group $U^{+}(t)$ over the parameter $t\in \mathbb{R}^{1}$
is a consequence of the $\ast$-weak continuity of
group \eqref{grG} of the Heisenberg equation \cite{DauL_5}.

To construct an infinitesimal generator of the group $\{U^{+}(t)\}_{t\in\mathbb{R}}$
we firstly differentiate the $nth$-term of expansion \eqref{krut_rozv}
in the sense of the point-by-point convergence of the space $\mathfrak{L}_{\gamma}$.
If $g\in\mathcal{D}(\mathcal{N})\subset\mathfrak{L}_{\gamma}(\mathcal{F}_\mathcal{H})$ similar to equality \eqref{Nint}
for $(1+n)th$-order dual cumulant \eqref{rozv_rec}, $n\geq1$, we derive
\begin{equation}\label{derivation3}
\begin{split}
       &\lim\limits_{t\rightarrow 0}\frac{1}{t}\mathfrak{A}_{1+n}^{+} \big(t,(Y\backslash X)_1,X \big) g_{s-n}(Y\backslash X)\psi_{s}=
       \sum\limits_{\mbox{\scriptsize $\begin{array}{c}{Z \subset Y\backslash X},\\{Z \neq\emptyset}\end{array}$}}
           \mathcal{N}_{\mathrm{int}}^{(|Z|+n)}(Z,X)g_{s-n}(Y\backslash X)\psi_{s}=\\
         &\sum\limits_{k=1}^{s-n}\frac{1}{k!}\sum\limits_{i_1\neq\ldots\neq i_{k}\in\{j_1,\ldots,j_{s-n}\}}
        \mathcal{N}_{\mathrm{int}}^{(k+n)}(i_1,\ldots,i_{k},X)g_{s-n}(Y\backslash X)\psi_{s}.
\end{split}
\end{equation}
Then according to equalities \eqref{infOper1} and \eqref{derivation3} for group \eqref{krut_rozv}  we obtain
\begin{equation*}
\begin{split}
   &\lim\limits_{t\rightarrow 0}\frac{1}{t}\big(\big(U^{+}(t)g\big) _{s}-g_{s}\big)\psi_{s}
     = \lim\limits_{t\rightarrow 0}\frac{1}{t}\big(\mathfrak{A}_{1}^+(t)g_{s}-g_{s}\big)\psi_{s}+\\
   &\sum_{n=1}^{s}\,\frac{1}{(s-n)!}\sum_{j_1\neq\ldots\neq j_{s-n}=1}^s\lim\limits_{t\rightarrow 0}
     \frac{1}{t}\mathfrak{A}_{1+n}^+ \big(t,(Y\backslash X)_1,X\big) g_{s-n}(Y\backslash X)\psi_{s}=\\
   &\mathcal{N}_{s}g_{s}\psi_{s}+
        \sum\limits_{n=1}^{s}\frac{1}{n!}
        \sum\limits_{k=n+1}^s \frac{1}{(k-n)!}\sum_{j_1\neq\ldots\neq j_{k}=1}^s
        \mathcal{N}_{\mathrm{int}}^{(k)}(j_1,\ldots,j_{k})
        g_{s-n}(Y\backslash\{j_1,\ldots,j_{n}\})\psi_{s},
\end{split}
\end{equation*}
where we used the identity
\begin{eqnarray}\label{id}
\sum\limits_{n=0}^{s}\frac{1}{n!}
        \sum_{j_1\neq\ldots\neq j_{n}=1}^s g_{s-n}(Y\backslash\{j_1,\ldots,j_{n}\})=
\sum\limits_{n=0}^{s}\frac{1}{n!}
        \sum_{j_1\neq\ldots\neq j_{n}=1}^s g_{n}(j_1,\ldots,j_{n})
\end{eqnarray}
which is valid in view of the Maxwell-Boltzmann statistics symmetry property.

Thus if $g\in\mathcal{D} (\mathfrak{B}^{+})\subset\mathfrak{L}_{\gamma}(\mathcal{F}_\mathcal{H})$
in the sense of the $\ast$-weak convergence of the space $\mathfrak{L}_{\gamma}(\mathcal{F}_\mathcal{H})$ we finally have
\begin{eqnarray*}
\mathrm{w^{\ast}-}\lim\limits_{t\rightarrow 0}\big(\frac{1}{t}\big(U^{+}(t)g - g\big)
 -\mathfrak{B}^{+}g\big)=0,
\end{eqnarray*}
where the generator  ${\mathfrak{B}}^{+}={\bigoplus\limits}_{n=0}^{\infty}
{\mathfrak{B}}^{+}_{n}$ of group \eqref{krut_rozv} is given by formula \eqref{d} (the dual operator to generator \eqref{1}).
\end{proof}

We now give examples of expansion \eqref{krut_rozv} and infinitesimal generator \eqref{d}.
The sequence $g=\big(0,g_{1}(1),0,\ldots \big)$ corresponds to the additive-type observable \cite{BG} and in this case
expansion \eqref{krut_rozv} for the group $U^{+}(t)$ get the form
\begin{eqnarray*}
       (U^{+}(t)g)_{s}(Y)= \mathfrak{A}_{s}^+(t,1,\ldots,s)
       \sum_{j=1}^s g_{1}(j), \quad s\geq 1.
\end{eqnarray*}
In the case of two-body interaction potential \eqref{H_Zag} operator \eqref{d} has the form
\begin{eqnarray}\label{dig}
        ({\mathfrak{B}}^{+} g)_{s}(Y) = \mathcal{N}_{s}(Y)g_{s}(Y)+
         \sum_{j_1\neq j_{2}=1}^s
         \mathcal{N}_{\mathrm{int}}^{(2)}(j_1,j_{2}) g_{s-1}(Y\backslash\{j_1\}), \quad s\geq 1,
\end{eqnarray}
where the operator $\mathcal{N}^{(2)}_{\mathrm{int}}$ is defined by formula \eqref{oper Nint2} for $n=2$.
If $\mathcal{H}=L^{2}(\mathbb{R}^3)$ in terms
of kernels of operators $g_{s}$, $s\geq 1$, for expression \eqref{dig} we have
\begin{eqnarray*}
        &&({\mathfrak{B}}^{+} g)_{s}(q_1,\ldots,q_s;q'_1,\ldots,q'_s)=\\
       &&-\frac{i}{\hbar}\Big(-\frac{\hbar^2}{2}\sum\limits_{i=1}^s(-\Delta_{q_i}+\Delta_{q'_i})+
       \sum\limits_{1=i<j}^s\big(\Phi^{(2)}(q'_i-q'_j)-\Phi^{(2)}(q_i-q_j)\big)\Big)
        g_s(q_1,\ldots,q_s;q'_1,\ldots,q'_s)-\\
        &&\frac{i}{\hbar}\sum\limits_{1=i\neq j}^s\big(\Phi^{(2)}(q'_i-q'_j)-\Phi^{(2)}(q_i-q_j)\big)g_{s-1}(q_1,\ldots,\E^{j},\ldots,q_s;q'_1,\ldots,\E^{j},\ldots,q'_s),
\end{eqnarray*}
where $(q_1,\ldots,\E^{j},\ldots,q_s)\equiv(q_1,\ldots,q_{j-1},q_{j+1},\ldots,q_s).$
This expression for a system of classical particles
is defined as a generator of the dual BBGKY hierarchy stated in \cite{CGP97},\cite{BG}.

In the paper \cite{BG} for classical systems of particles
an equivalent representation of group \eqref{krut_rozv} was used, namely for the case under consideration
group the $U^{+}(t)$ has the following representation
\begin{eqnarray}\label{rds}
U^{+}(t)=\mathcal{G}(t)+ \sum\limits_{n=1}^{\infty}\frac{1}{n!}
\big[\ldots\big[\mathcal{G}(t),\underbrace{\mathfrak{a}^{+} \big],\ldots,\mathfrak{a}^{+}}_{\hbox{n-times}}\big]=
 e^{-\mathfrak{a}^{+}}\mathcal{G}(t)e^{\mathfrak{a}^{+}},
\end{eqnarray}
where $\big[\, .\, , \,.\,\big]$ is a commutator, the operator $\mathfrak{a}^+$ (an analog of the creation
operator) is defined on the space $\mathfrak{L}_{\gamma}(\mathcal{F}_\mathcal{H})$ by the formula
\begin{eqnarray}\label{oper_znuw}
         \big( \mathfrak{a}^+ g \big)_{s}(Y):=\sum_{j=1}^s \, g_{s-1}(Y \backslash \{j \}),
\end{eqnarray}
the operators $e^{\pm \mathfrak{a}^{+}}$ are defined on the space $\mathfrak{L}_{\gamma}(\mathcal{F}_\mathcal{H})$
by the expansions
\begin{eqnarray*}
\big(e^{\pm \mathfrak{a}^{+}}g\big)_s(Y)=\sum_{n=0}^s\,\frac{(\pm 1)^n}{n!}\sum_{j_1\neq\ldots\neq j_{n}=1}^s
            g_{s-n}\big(Y\backslash \{j_1,\ldots,j_{n}\}\big),  \quad\!\! s\geq 1.
\end{eqnarray*}
and the group of operators $\mathcal{G}(t)$ is defined by expression \eqref{grG}.

Representation \eqref{rds} is true in consequence of definition \eqref{oper_znuw} of the operator $\mathfrak{a}^{+}$
and the validity of the equality
\begin{equation}\label{rrr}
\begin{split}
&\big(\frac{1}{n!}\big[\ldots\big[\mathcal{G}(t),\underbrace{\mathfrak{a}^{+} \big],\ldots,\mathfrak{a}^{+}}_{\hbox{n-times}}\big]g\big)_s(Y)=\\
&\frac{1}{(s-n)!}\sum_{j_1\neq\ldots\neq j_{s-n}=1}^s
      \mathfrak{A}_{1+n}^{+} \big(t,(Y\backslash X)_1,X\big) \,
      g_{s-n}(Y\backslash X),  \quad\!\! s\geq 1,
\end{split}
\end{equation}
where $X\equiv Y\backslash\{j_1,\ldots,j_{s-n}\}$ and $Y\backslash X=(j_1,\ldots,j_{s-n})$.
For example, if $n=1$ we have
\begin{equation*}
\begin{split}
&\big(\big[\mathcal{G}(t), \mathfrak{a}^{+} \big]g\big)_s(Y)=\sum_{j=1}^s\big(\mathcal{G}_s(t,Y) -
\mathcal{G}_{s-1}(t,Y\backslash \{j\})\big)g_{s-1}(Y\backslash \{j\})= \\
&\sum_{j=1}^s \mathfrak{A}_{2}^{+}\big(t,(Y\backslash \{j\})_1,j \big) \,g_{s-1}(Y\backslash \{j\})
\end{split}
\end{equation*}
and since identity \eqref{id} holds we obtain equality \eqref{rrr}.

We can obtain one more representation of the group $U^{+}(t)$, if we express the dual cumulants $
\mathfrak{A}_{1+n}^+(t), ~n\geq1,$ of groups \eqref{grG} with respect to the $1st$-order and $2nd$-order
dual cumulants. In fact it holds
\begin{equation*}
    \mathfrak{A}_{1+n}^{+}\big(t,(Y\backslash X)_1, X \big)=
    \sum\limits_{\substack{Z\subset X,\\Z\neq \emptyset}} \mathfrak{A}_{2}^{+}\big(t,Y\backslash X,Z\big)
    \sum\limits_{\mathrm{P}:\,X \backslash Z ={\bigcup\limits}_i X_i}
    (-1)^{|\mathrm{P}|}\,|\mathrm{P}| ! \,\, {\prod}_{i=1}^{|\mathrm{P}|} \mathfrak{A}_{1}^{+}\big(t,X_{i} \big),
\end{equation*}
where ${\sum\limits}_{\substack{Z\subset X,\\Z\neq \emptyset}}$ is a sum over all nonempty subsets $Z\subset X$ of the set $X$.
Then taking into account an identity
\begin{equation*}
  \sum\limits_{\mathrm{P}:\,X \backslash Z ={\bigcup\limits}_i X_i}
    (-1)^{|\mathrm{P}|}\,|\mathrm{P}| ! \,\, {\prod}_{i=1}^{|\mathrm{P}|} \mathfrak{A}_{1}^{+}\big(t,X_{i} \big)g_{s-n}(Y\backslash X)=
  \sum\limits_{\mathrm{P}:\,X \backslash Z ={\bigcup\limits}_i X_i}
    (-1)^{|\mathrm{P}|}\,|\mathrm{P}| ! \,g_{s-n}(Y\backslash X)
\end{equation*}
and equality \eqref{eq1} we get the following representation of the group $U^{+}(t)$
\begin{equation*}
\begin{split}
    &\big(U^{+}(t)g\big) _{s}(Y)=\mathfrak{A}_{1}^{+}\big(t,Y\big)g_{s}(Y)+\\
    &\sum_{n=1}^s\,\frac{1}{(s-n)!}\,\sum_{j_1\neq\ldots\neq j_{s-n}=1}^s\,\,\,
      \sum\limits_{\substack{Z\subset X,\\Z\neq \emptyset}}\,(-1)^{|X\backslash Z|}\,\, \mathfrak{A}_{2}^{+}\big(t,Y\backslash X,Z\big) \,
      g_{s-n}(Y\backslash X).
\end{split}
\end{equation*}

An infinitesimal generator of group \eqref{rds} has the following form
\begin{eqnarray}\label{rdg}
{\mathfrak{B}}^{+}=\mathcal{N}+ \sum\limits_{n=1}^{\infty}\frac{1}{n!}
\big[\ldots\big[\mathcal{N},\underbrace{\mathfrak{a}^{+} \big],\ldots,\mathfrak{a}^{+}}_{\hbox{n-times}}\big]=
e^{-\mathfrak{a}^{+}}\mathcal{N}e^{\mathfrak{a}^{+}}.
\end{eqnarray}
Representation \eqref{rdg} is true in consequence of definition \eqref{oper_znuw} of the operator $\mathfrak{a}^{+}$
and the validity of an identity
\begin{equation*}
\big(\big[\ldots\big[\mathcal{N},\underbrace{\mathfrak{a}^{+} \big],\ldots,\mathfrak{a}^{+}}_{\hbox{n-times}}\big]g\big)_s(Y)=
\sum\limits_{k=n+1}^s \frac{1}{(k-n)!}\sum\limits_{j_1\neq\ldots\neq j_{k}=1}^s
        \mathcal{N}_{\mathrm{int}}^{(k)}(j_1,\ldots,j_{k})g_{s-n}\big(Y\backslash\{j_1,\ldots,j_{n}\}\big),
\end{equation*}
which in the case of a two-body interaction potential reduces to the following one
\begin{eqnarray*}
\big(\big[\mathcal{N},\mathfrak{a}^{+}\big]g\big)_s(Y)=
\sum\limits_{j_1\neq j_{2}=1}^{s} \mathcal{N}_{\mathrm{int}}^{(2)}(j_1,j_{2})g_{s-1}\big(Y\backslash\{j_1\}\big).
\end{eqnarray*}
%\end{remark}

\section{Conclusion}
\setcounter{equation}{0}

The concept of cumulants \eqref{cumulant} of groups \eqref{groupG} of the von Neumann equations
forms the basis of group expansions for quantum evolution equations, namely, the von Neumann hierarchy
for correlation operators \cite{GerSh}, as well as the BBGKY hierarchy for $s$-particle density operators \cite{GerS}
and the dual BBGKY hierarchy \cite{GerR}. In the case of quantum systems of particles
obeying Fermi or Bose statistics groups \eqref{groupKum},\eqref{RozvBBGKY} and \eqref{krut_rozv} have different structures.
The analysis of these cases will be given in a separate paper.

We have stated the properties of groups \eqref{groupKum} and \eqref{RozvBBGKY}
on the space $\mathfrak{L}^{1}_{\alpha}(\mathcal{F}_\mathcal{H})$
and dual group \eqref{krut_rozv} on $\mathfrak{L}_{\gamma}(\mathcal{F}_\mathcal{H})$.
To describe the evolution of infinitely many particles \cite{CGP97} it is necessary to define the one-parametric
family of operators  \eqref{RozvBBGKY} on more general spaces than  $\mathfrak{L}^{1}_{\alpha}(\mathcal{F}_\mathcal{H})$, for example,
on the space of sequences of bounded operators containing the equilibrium states \cite{Gen}. For dual group \eqref{krut_rozv}
the problem lies in the definition of functional \eqref{averageD} for operators from the corresponding spaces.
In both these cases every term of the corresponding expansions contains the divergent traces \cite{CGP97},\cite{GerS},\cite{BG}
and the analysis of such a question for quantum systems remains an open problem.

On the space $\mathfrak{L}_{\gamma}(\mathcal{F}_\mathcal{H})$ one-parametric mapping \eqref{krut_rozv} is not a strong continuous group.
The group $\{U^{+}(t)\}_{t\in\mathbb{R}}$ of operators \eqref{krut_rozv} defined on the space $\mathfrak{L}_{\gamma}(\mathcal{F}_\mathcal{H})$
is dual to the strong continuous group $\{U(t)\}_{t\in\mathbb{R}}$ of operators \eqref{RozvBBGKY} for the BBGKY hierarchy defined on
the space $\mathfrak{L}_{\alpha}^{1}(\mathcal{F} _\mathcal{H})$ and the fact that one is a $C_{0}^{\ast}$-group
follows also from general theorems about properties of the dual semigroups \cite{BR},\cite{Pazy}.

As mention above the group $\{\mathcal{G}(-t)\}_{t\in\mathbb{R}}$ of operators \eqref{groupG} preserves positivity \cite{BanArl},\cite{AL}.
The same property must be valid for the group $\{U(t)\}_{t\in\mathbb{R}}$ of operators \eqref{RozvBBGKY}
for the BBGKY hierarchy, but how to prove this property one is an open problem.

We have constructed infinitesimal generators \eqref{Nnl}, \eqref{1} on the subspace $\mathfrak{L}^{1}_{\alpha, 0}\subset\mathfrak{L}^{1}_{\alpha}(\mathcal{F}_\mathcal{H})$ and generator \eqref{d} on $\mathcal{D}(\mathfrak{B^+})\subset\mathfrak{L}_{\gamma}(\mathcal{F}_\mathcal{H})$. The question of how
to define the domains of the definition
$\mathcal{D}(\mathfrak{N}), \mathcal{D}(\mathfrak{B})$ and $\mathcal{D}(\mathfrak{B^+})$
of corresponding generators \eqref{Nnl}, \eqref{1} and \eqref{d} is an open problem \cite{BR},\cite{AL}.

\end{document}